\documentclass{ab}

\usepackage{graphicx}
\usepackage{epstopdf}
\usepackage{natbib}

\begin{document}

\title{ADAM Low- and Medium-Resolution Spectrograph for 1.6-m AZT-33IK  Telescope}

\author{V.L.~Afanasiev$^1$, S.N.~Dodonov$^1$, V.R.~Amirkhanyan$^{1,2}$,A.V. Moiseev$^1$}

\institute{$^1$Special Astrophysical Observatory, Russian Academy of Sciences, Nizhnij Arkhyz, 369167, Russia
\\
$^2$Sternberg Astronomical Institute, M.V.Lomonosov Moscow State University, Moscow, 119992 Russia
}
 \titlerunning{ADAM Low- and Medium-Resolution Spectrograph}

\authorrunning{Afanasiev et al.}

\date{July 21, 2016 / Revised: September 5, 2016}
\offprints{V. Afanasiev  \email{vafan@sao.ru}, A. Moiseev  \email{moisav@sao.ru} }

\abstract{
We describe the design of a  low- and medium-resolution
spectrograph (\mbox {$R\approx 300$--$1300$}) developed at
the Special Astrophysical Observatory of the Russian Academy of
Sciences (SAO RAS) for the 1.6-m \mbox {AZT-33IK} telescope of Sayan
Observatory of the Institute of Solar-Terrestrial Physics of the
Siberian Branch of the Russian Academy of Sciences. We report the
results of laboratory measurements of the parameters of the
instrument and tests performed on the SAO RAS 1-m Zeiss-1000 telescope.
We measured the total quantum efficiency of the  ``spectrograph +
telescope + detector'' system on AZT-33IK telescope, which at its
maximum reaches 56\%. Such a hight transparency of the spectrograph
allows it  to be used with the 1.6-m telescope to determine the types
and redshifts of objects with  magnitudes $m_{\rm
AB}\approx20$--$21$, that  was confirmed by actual
observations.

}

\maketitle

\section{Introduction}
\label{intro}

The capabilities of the spectroscopy of faint galaxies and stars on
moderate-size telescopes with aperture diameters of 1--1.5~m are
determined by the efficiency of the spectrographs employed. Despite
their excellent optical quality the well-known universal commercial
spectrographs manufactured by
Boller~\&~Chivens~\citep{mack} and Carl Zeiss
Jena~\citep{gey} have rather low transmission
(15--20\%). In the last three decades  focal reducers based on  refractive  optics exclusively became very popular on large
telescopes. The prototype of such devices is EFOSC camera of the
3.6-m ESO telescope~\citep{buzoni}. At the end of
the last century an efficient low- and medium-resolution DFOSC
--- Danish Faint Object Spectrograph and
Camera~\citep{ander} spectrograph equipped with a
focal reducer and similar to EFOSC  was developed for the
1.54-m telescope at Copenhagen Observatory. The
transmission of the optics of such a spectrograph exceeds 70\%, and
the total quantum efficiency of the ``spectrograph + telescope +
CCD'' system is of about 30\%. Since then nine such spectrographs
have been made for small telescopes of various observatories. Note
that the spectrographs mentioned above are multimode devices, which,
in our opinion, is a justified solution in observations on large
telescopes, where the observing process is strictly regulated.
Multimode spectrographs have many optical elements, which reduce
their efficiency and make their use on small telescopes unfeasible.
It is more efficient to dedicate small telescopes with highly specialized  equipment to single task. Currently,
modern optical elements (new glass types, AR coatings, volume
holographic gratings, etc.) and readily available commercial systems
with  highly efficient CCD  make it in
principle possible to make a low-resolution spectrograph with the
efficiency greater than \mbox {50--60}\%.

In this paper we describe a low- and medium-resolution spectrograph (from
$R=\lambda/\delta\lambda\approx300$ to $R\approx1300$) designed for the 1.6-m
AZT-33IK telescope of Sayan Observatory of the Institute of Solar-Terrestrial
Physics of the Siberian Division of  the Russian Academy of  Sciences.
The device is intended for acquiring optical spectra of faint
galaxies and stars within the framework of the program of ground-based support
of ``Spectrum--Rontgen--Gamma'' space observatory.

\begin{figure*}
\centerline{\includegraphics[scale=0.6]{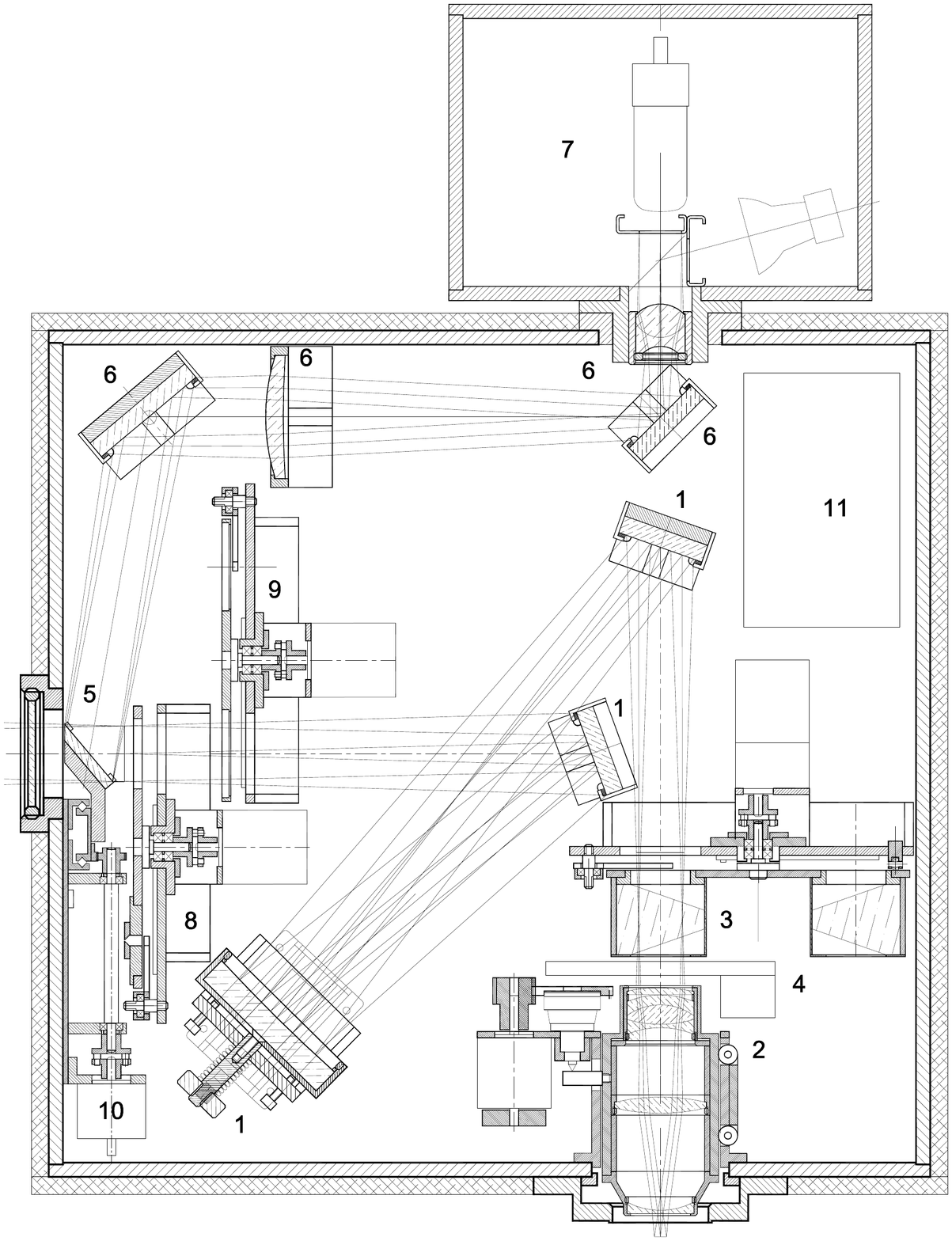}}
\caption{Optomechanical layout of ADAM spectrograph: ({\it
1})---collimator consisting of an off-axis spherical mirror and two
beam bending flat mirrors; ({\it 2})---camera; ({\it
3})---the grisms wheel; ({\it 4})---electromechanical
shutter; ({\it 5})---calibration unit mirror that can be introduced
into the beam; ({\it 6})---projector of the calibration unit
consisting of a two-lens condenser, a field lens, and two
beam bending flat mirrors; ({\it 7})---calibration lamps unit; ({\it 8})---the slits wheel; ({\it
9})---the filters wheel; ({\it 10})---the mechanism for
introducing/withdrawing the calibration mirror; ({\it 11})---the
control computer.} \label{fig_optics}
\label{fig1}
\end{figure*}

\section{Design of the instrument}

When designing the spectrograph we took advantage of the experience
obtained in  the process of  the development of
SCORPIO~\citep{AfanasievMoiseev2005} and \mbox
{SCORPIO-2}~\citep{AfanasievMoiseev2011} instruments
for the SAO RAS \mbox{6-m} telescope, which are similar to  EFOSC and which
are remote-control oriented. In these instruments we for the first
time in the practice of Russian astronomical instrument design used
volume phased holographic gratings~\citep{barden},
which are 30--80\% more efficient  than usual ruled gratings.

The spectrograph that we developed is mounted on a controlled
rotation stage in the Ritchey--Chretien focus (with the equivalent
focal length of 30~m) of the 1.6-m  \mbox {AZT-33IK} telescope. The
instrument must have high transmission  which,
combined with a modern Deep Depletion \mbox {CCD detector}, will make
it possible to acquire spectra of faint (down to $21^{\rm m}$)
starlike targets in the blue and red parts of the spectrum within
reasonable exposure times. The spectrograph can also operate in the
direct imaging mode to ensure accurate positioning of the target on the slit.

\begin{table}
\caption{Principal parameters of ADAM spectrograph mounted on AZT-33IK telescope} \label{table_main}
\medskip
\begin{tabular}{l|c}
\hline
Equivalent focal ratio  &  $F/4.1$ \\
Field of view    &  $3\farcm46\times3\farcm46$ \\
Image scale& $0\farcs81\,\mbox{px}^{-1}$\\
Wavelength range& $3600$--$10\,000$~\AA\\
QE$_{\rm max}^*$ (telescope +&\\
spectrograph + CCD) & 56\% \\
Spectral resolution& $FWHM=6$--$15$~\AA\\
(for the $1\farcs5$-wide slit)&  ($R=1320$--$270$)  \\
\hline
\end{tabular}

\small $^*$~--- maximum quantum efficiency\\
\end{table}

Note also that one of the crucial requirements to the spectrograph is
that it should be capable  of smooth operation at ambient
temperatures ranging from $-30\degr$C  to \mbox {$+20\degr$C} as
determined by the climatic conditions of Sayan Observatory.

\subsection{Optical Layout}

The spectrograph is assembled in accordance with the traditional
scheme (Fig.~\ref{fig_optics}): an off-axis mirror
collimator forming the exit pupil in the parallel beam and a
catadioptric lens camera. The collimator has the focal length of
500~mm and produces a 27-mm diameter exit pupil. Two additional
flat beam bending mirrors are used to reduce the overall size of the
instrument. The spectrograph camera consists of a three-component
six-lens apochromat with a focal length of 109~mm with a geometric
focal ratio of  $F/3$. The physical focal ratio---the ratio of the
focal length of the camera to the diameter of the collimated
beam---is equal to $F/4.36$. Lenses have seven-layer antireflection
coating applied to them, which operate in the  0.36--1~$\mu$m
wavelength interval and have a transmission coefficient of no less
than 98\%. Mirrors are coated with a reflective protected silver
layer with a reflection coefficient of 99\% throughout the entire
operating wavelength interval. Antireflection and reflection coatings
have been computed and applied at   Opto-Technological
Laboratory LTD\footnote{\tt www.optotl.ru}. The mechanical and optical
parts of the spectrograph were manufactured at mechanical  workshops
of the Special Astrophysical Observatory of the Russian Academy of
Sciences (SAO RAS). The detector employed is the NEWTON CCD system described
below in Section~\ref{sec_ccd}.
Table~\ref{table_main} lists the principal parameters
of the spectrograph.

\begin{figure*} 
\centerline{\includegraphics[scale=0.65]{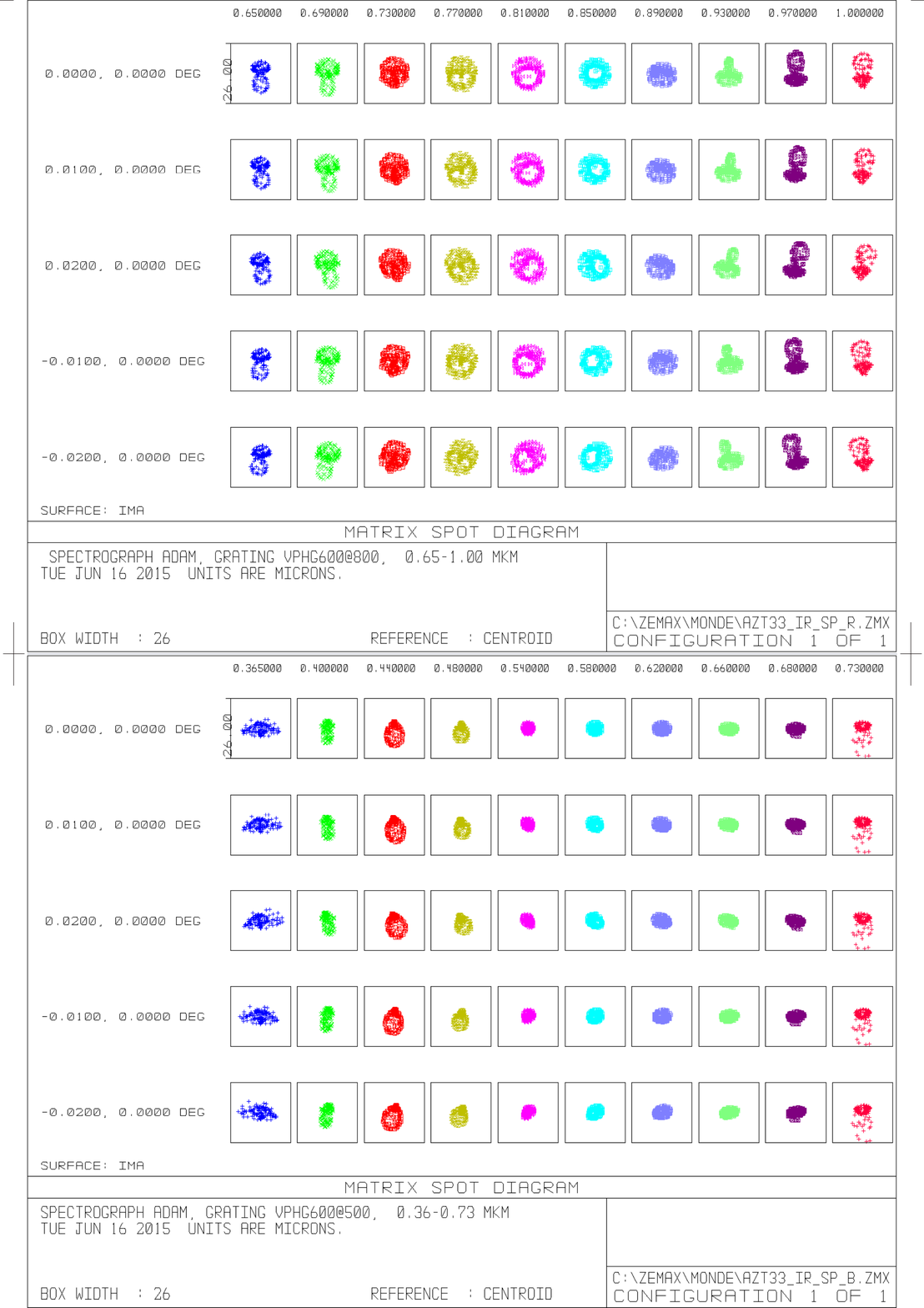}}
\caption{Spot diagrams in the spectroscopy mode. The top
panel: the red part of the spectrum,  0.65--1.00~$\mu$m, the bottom
panel: the blue part of the  spectrum, 0.365--0.73~$\mu$m. The square
side size is 26~$\mu$m, which corresponds to the  1~pixel of the CCD.} \label{PSF}
\end{figure*}

The equivalent focal length of the  ``telescope + spectrograph''
system is  6584~mm, which corresponds to the image scale of
32~$\mu$m\,arcsec$^{-1}$ in the detector plane. The unvignetted field
of view has the size of \mbox {$3.5\times3.5$} arcmin. The
spectrograph uses volume holographic gratings as the dispersing
elements. The parameters of these gratings are given below. The size
of the circle of diffusion in the spectrograph does not exceed
15~$\mu$m in the \mbox {0.35--1.0}~$\mu$m wavelength interval.
Figure~\ref{PSF} shows the computed spot diagrams
for two spectral intervals: the red and the blue.

The telecentric optical layout of the calibration unit consists of a
two-lens condenser and a field lens, which produce images of the area
illuminated by the calibration lamps at the exit pupil of the
collimator.

\subsection{Mechanical Design of the Spectrograph}

The device is made in the form of a rigid duralumin case with several units
attached to it:

\begin{list}{}{
\setlength\leftmargin{2mm} \setlength\topsep{2mm}
\setlength\parsep{0mm} \setlength\itemsep{2mm} }
 \item[$\bullet$]  CCD detector;
 \item[$\bullet$]  Calibration lamps unit;
 \item[$\bullet$]  Mechanism for introducing/withdrawing the diagonal mirror in front of the slit;
 \item[$\bullet$]  Slits wheel;
 \item[$\bullet$]  Filters wheel;
 \item[$\bullet$]  Grisms wheel;
 \item[$\bullet$]  Adjustable spherical collimator mirror;
 \item[$\bullet$]  Four flat mirrors and the field lens of the calibration unit;
 \item[$\bullet$]  Spectrograph camera focusing mechanism.
\end{list}

\begin{figure*} 
\centerline{\includegraphics[scale=0.6]{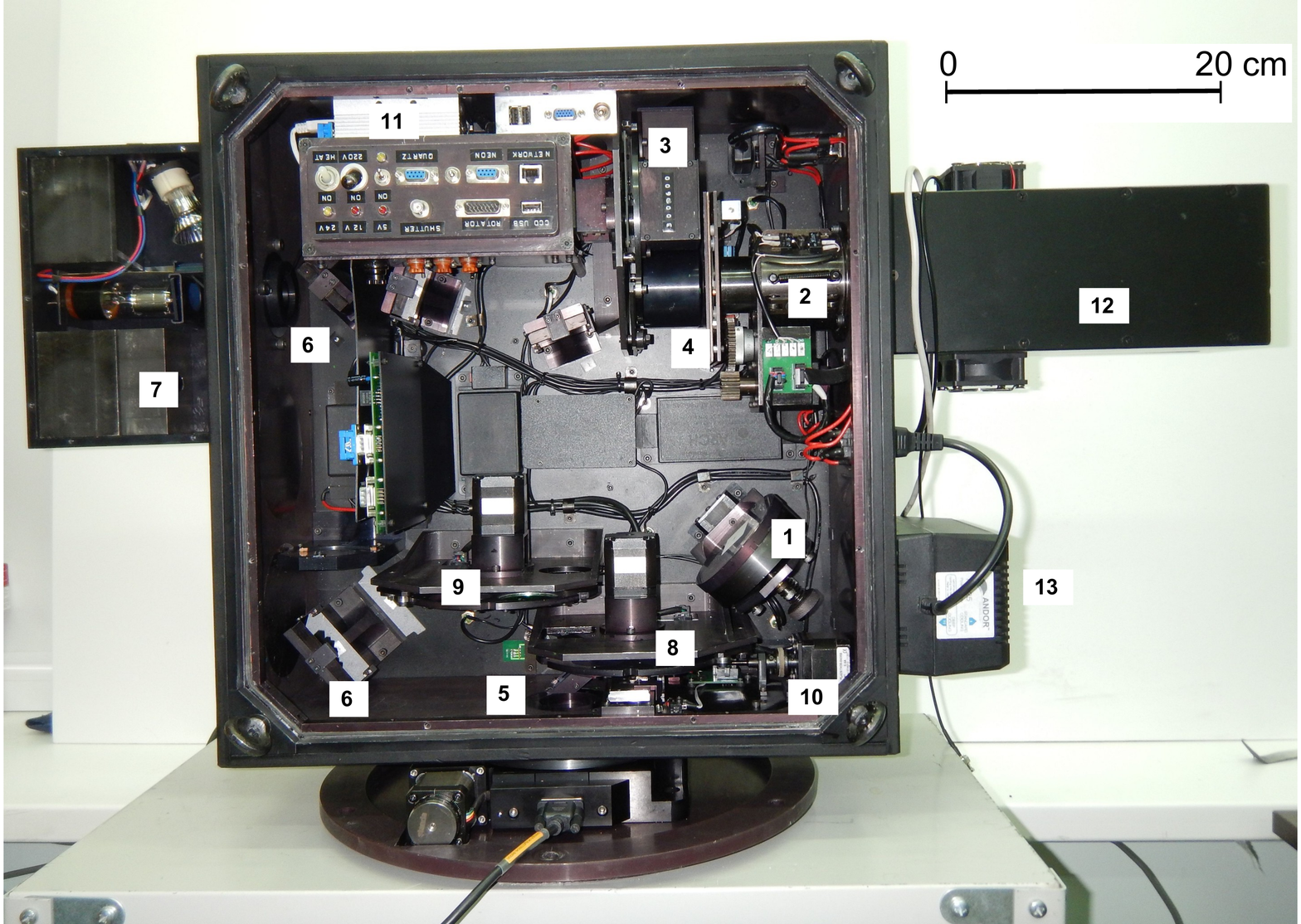}}
\caption{ADAM spectrograph with removed cover: ({\it 1})---collimator; ({\it 2})---camera; ({\it 3})--- the grisms wheel; ({\it
4})---electromechanical shutter; ({\it 5})---calibration mirror; ({\it 6})---projector of the
calibration unit; ({\it 7})---calibration lamps unit; ({\it 8})---the slits wheel; ({\it
9})---the filters wheel; ({\it 10})---mechanism for
introducing/withdrawing of  the calibration mirror; ({\it
11})---control computer; ({\it 12})---CCD camera; ({\it 13})---CCD
power supply unit.} \label{fig_inside}
\end{figure*}

The following devices are also mounted inside the spectrograph:
power supply sources, spectrograph control board, thermostat board,
control computer, and spectrograph control panel.
One of the spectrograph walls is removable to provide access to all
optomechanical elements for adjustment and cleaning of optics.
Figure~\ref{fig_inside} shows the layout of the elements inside
the spectrograph. The weight of the spectrograph without the
rotation stage and detector is 39~kg.

\subsection{CCD Detector}
\label{sec_ccd} 

As a detector ADAM uses NEWTON system
of highly sensitive optical signal registration device manufactured
by  ANDOR\footnote{\tt http://www.andor.com/} (Northern Ireland)
based on a E2V CCD30-11 detector. The camera is connected to the
control computer via  \mbox{USB-2} interface.
Table~\ref{table_ccd} lists the principal parameters
of the device according  to the manufacturer's specification sheet.
The spectrograph control software allows setting the working
temperature of the CCD, choose modes with different gains, readout
speeds and noise. The data are stored as a standard
16-bit FITS files.

\begin{table}
\caption{Parameters of  E2V CCD30-11 CCD} \label{table_ccd}
\medskip

\begin{tabular}{l|l}
\hline
 Type     &  Thin, back illuminated\\ 
 Format & $1024\times256$ \\
Pixel size &  $26\mu\times26\mu$ \\
QE$_{\rm max}$ & 95\%\\
RON$_{\rm min}^{*}$  & 3.5  $e^-$  ($T=100\degr$~C)\\
Dark current & $<0.3e^-$~px$^{-1}$\,min.$^{-1}$\\
Gain, $e^-$\,ADU$^{-1}$& 2.6 (high)\\
              &5.2 (normal)\\
              &11 (low) \\
\hline
\end{tabular}

\small $^*$~--- minimal readout noise
\end{table}

The three-stage Peltier cooling system of the CCD camera provides a
working temperature of $-100\degr$C. Heat from the hot Peltier
junction is removed by a liquid cooling system consisting of CW-3000
powerful commercial chiller and BC103 climate-control unit. The
climate-control unit keeps the coolant (propylene glycol) temperature
at a constant level. The working temperature of the coolant is
$4$--$5\degr$C, and it completely prevents any condensation effects
on hoses used to transport the coolant liquid from the chiller to the
CCD camera.

\subsection{Calibration Unit}

The output area of the calibration path is illuminated by two
calibrating lamps: (1) a Ne-Ar-He filled lamp (hereafter referred to
as the NEON lamp) with line spectrum to calibrate the wavelength
scale and (2) continuum-spectrum lamp to produce the  ``flat  field''
(hereafter referred to as the FLAT lamp). A combination of SZS7 and
SS1 filters is used to equalize the brightness of lines in the red
and blue parts of the spectrum of the NEON lamp. The FLAT lamp is
equipped with an SZS7 glass filter, which reduces the flux from the
lamp at wavelengths longer than 5500~\AA, which is necessary for
generating more uniform detector illumination in the spectroscopic
mode.

\subsection{Slits}

Wheel~{\it 1} contains holders with five slits for spectroscopic
observations. The slit widths are strictly fixed and correspond to
$1''$, $1\farcs5$, $2''$, $3''$, and  $10''$ sizes in the focal
plane.

\subsection{Filters}

Wheel~{\it 2} allows mounting of five filters of diameter 50~mm and
 to 4~mm thickness. In one of the filter slots the OS11 filter is located,
which cuts off the second order in observations with
VPHG600R grism.

\subsection{Diffraction Gratings}

Wheel~{\it 3}  contains mounts with three  grisms
(combination of a transparent diffraction grating and two prisms).
The spectrograph uses volume phase holographic gratings manufactured
by Wasatch Photonics\footnote{\tt http://wasatchphotonics.com/}. The
parameters of the grisms  are listed in Table~\ref{tab_grism}.

\begin{table*}
\caption{Parameters of grisms } 
\label{tab_grism}
\medskip
\begin{tabular}{l|c|c|c|c}
\hline
Name& lines/mm& wavelength range & dispersion & sp. resolution\\
                              &                  &               \AA                  &           \AA\,px$^{-1}$                      & ($1\farcs5$-slit)\\
\hline
VPHG300  &  300 & ~~3510--10300 &  6.4--7.6   & 273--678     \\
VPHG600G &  600 & 3590--7250    &  3.2--3.7   & 561--980  \\
VPHG600R &  600 & ~~6430--10030 &  3.2--3.6   & 1005--1319   \\
 \hline
\end{tabular}
\end{table*}

\subsection{Thermal Stabilization System}

The spectrograph case is covered with heat-insulating material. The
case of the suspended part of the spectrograph ensures thermal
stabilization of the instrument at a level above $+5\degr$C at
ambient temperatures below 0$\degr$C. Eight 2T907A transistors
mounted on the base plate of the spectrograph case are used as
heating elements. The bridge connection with a termistor-based
temperature sensor is set so that a decrease of the case temperature
below the given threshold causes the increase of the collector
current of the transistors. The thermal stabilization system keeps
the plate temperature constant to within several tenths of degree.
The maximum power dissipated by the thermostat does not exceed 60 W.

\subsection{Rotation Stage}

The spectrograph is mounted on a Rotation Stage (Standa\footnote{\tt
http://www.standa.lt/}, 8MR190-90-4247), which allows rotating the
entire instrument about its axis in order to change position angle of the slit on the
sky. The stage can be turned with a relative accuracy of about
$0\fdg015$, which corresponds to one motor step. The zero point
position is controlled by a Hall effect sensor. To prevent   cable
kinking, the control program allows the rotator  to be turned only
within $\pm90\degr$ relative to the zero position.

\begin{figure*}
\centerline{\includegraphics[scale=0.9]{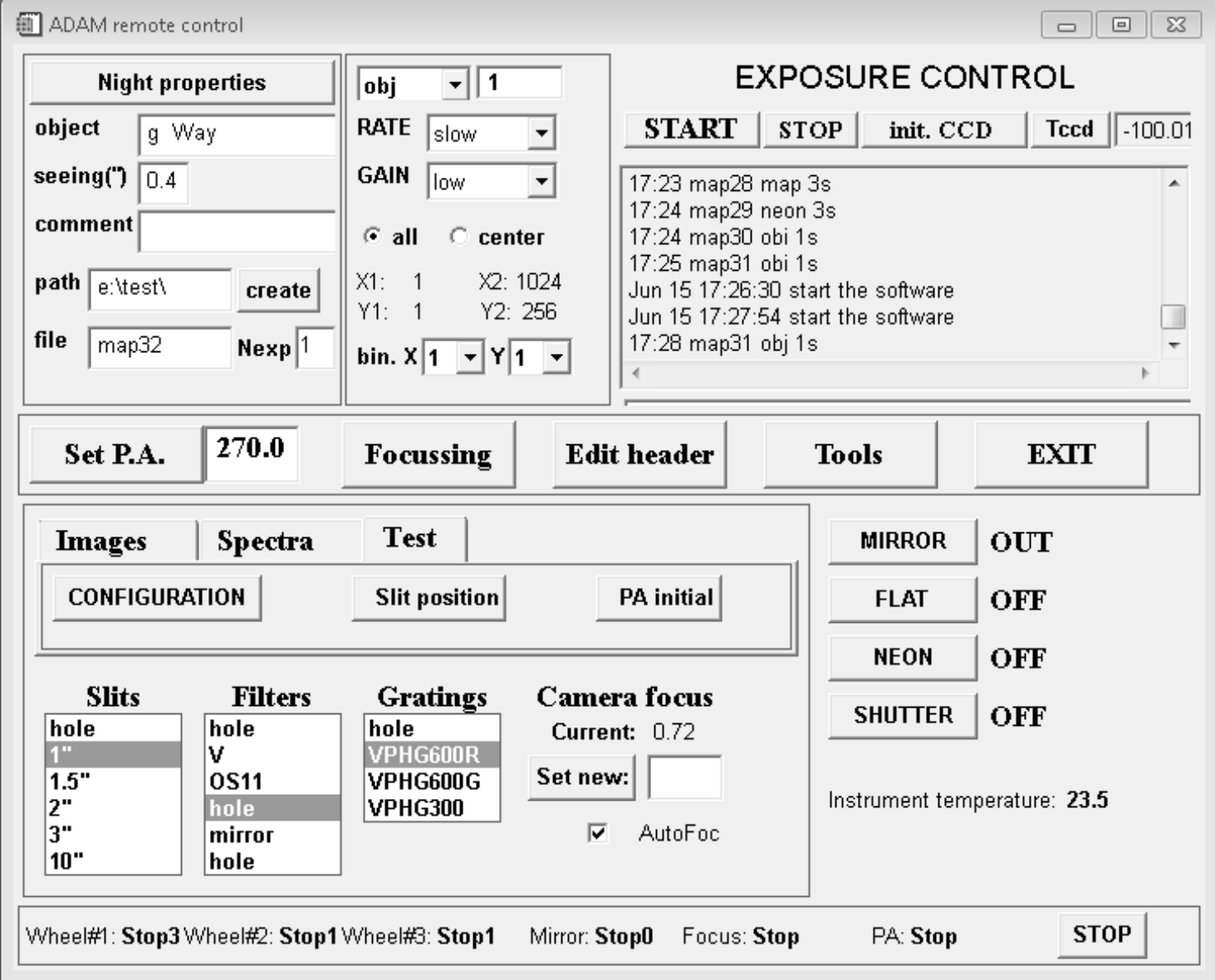}}
\caption{Spectrograph control menu.} \label{fig_interface}
\end{figure*}

\section{Control system}

\subsection{Control Electronics}
The process of observation includes the transition from direct image
mode to the spectroscopic mode, change of filters, grisms,
switching on and off calibration lamps, etc. This is what the built-in control
system of the spectrograph does. Its ``brain'' is an  ARK-1122F
industrial computer manufactured by Advantech. Computer commands are
received by Atmega8535 microprocessor, which transmits them to
actuator mechanisms via power control elements. Here is the list of
such mechanisms in our case:
\begin{list}{}{
\setlength\leftmargin{2mm} \setlength\topsep{2mm}
\setlength\parsep{0mm} \setlength\itemsep{2mm} }
 \item (1) shutter;
 \item (2)  slits  wheel ({\it 1}) with six positions;
 \item (3) filters  wheel({\it 2}) with six positions;
 \item (4) grids wheel with four positions;
 \item (5) mechanism for introducing/ withdrawing the calibration illumination mirror;
 \item (6) focus of the camera;
 \item (7) rotating stage;
 \item (8) calibrating FLAT lamp;
 \item (9) calibrating NEON lamp.
\end{list}
The shutter is actuated by a solenoid; mechanisms  2 to 7---by step
motors; calibrating lamps---by application of high voltage. The
built-in microprocessor program actuates the mechanisms, and
also determines the logic of their operation and controls their
state. To facilitate the assembly and increase the repairability of
actuator mechanisms, they have local boards mounted to accommodate
some of the elements of the control system. These include aspect
sensors, electronic keys, connectors. The temperature of the bearing
plate of the spectrograph is controlled by LM75 sensor.

The above technical solutions made it possible to substantially reduce
external communications and intelligent computer load, and simplify changing of the
spectrograph operation mode.

\subsection{Control Software}

ARK-1122F computer with the software that controls the spectrograph,
its state, and data acquisition, works under Windows 7 OS. The control
interface is written in  IDL\footnote{\tt
http://www.harrisgeospatial.com/} and allows sending commands to both
the local microprocessor and the CCD server (programs written by
T.~A.~Fatkhullin, SAO RAS). Fig.~\ref{fig_interface} shows the
main menu of exposure and spectrograph control. The observer may
reside far away from the telescope and control the instrument via
TCP/IP protocol using NetOp\footnote{\tt http://www.netop.ru/} remote
computer administration suite.

All parameters  of the current state of the spectrograph (position of
movable elements, temperature, etc.), and the main parameters of the telescope
(coordinates, focus) are automatically written to the FITS headers.

\begin{figure}
\centerline{\includegraphics[scale=0.8]{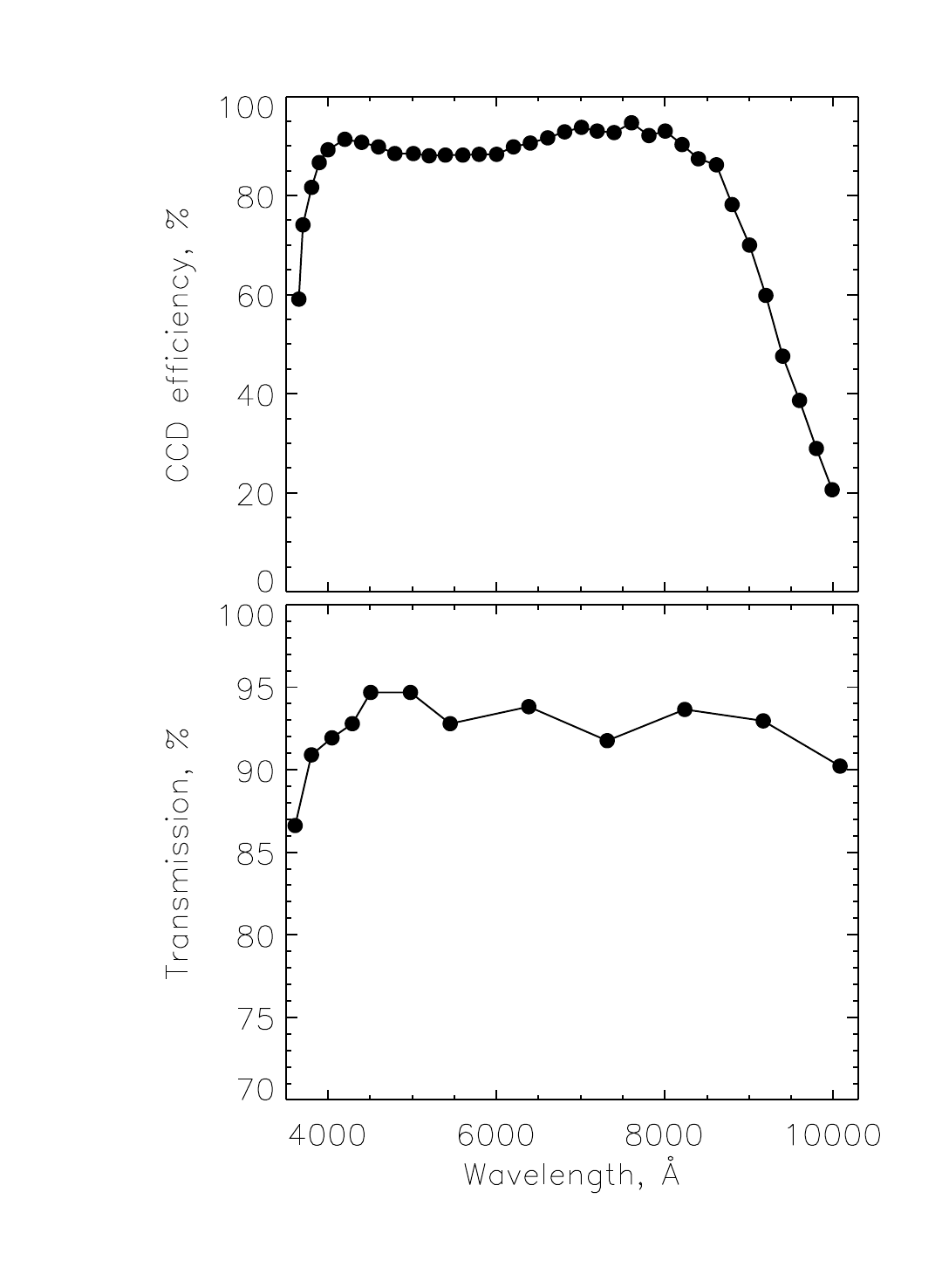} }
\caption{Results of laboratory measurements of the quantum efficiency of the CCD
(the top panel) and transmission of the spectrograph optics without grisms
(the bottom panel).}
\label{fig_trans}
\end{figure}

\section{Results of tests}

\subsection{Laboratory Tests}

After making the spectrograph we conducted laboratory measurements
of the critical operation parameters such as the transmission of the
optics, quantum efficiency, etc.

\subsubsection{Quantum Efficiency of the CCD}

We performed our measurements using MDR-41 monochromator in the
0.36--1.0~$\mu$m wavelength range. We used a halogen lamp as the
source of light and an Optronic Laboratory rated silicon LED as the
detector. The flux from the LED was recorded by OL730D radiometer of
the same manufacturer. Our measurements showed that the quantum
efficiency agrees with the nominal ratings of  E2V CCD30-11 detector:
it was greater than 90\% in the 0.42--0.83~$\mu$m wavelength interval
and more  than 40\% at the boundaries of the measurement range
(Fig.~\ref{fig_trans}).

\subsubsection{Transmission of  the Optics}

To determine the transmission of the optics without dispersing elements,
we compared the monochromatic flux at the entrance and output of the spectrograph
in the 0.36--1.0~$\mu$m wavelength interval. The monochromatic source with
a relative aperture of $F/20$ was formed using Jarrel Ash monochromator
and a halogen lamp. The measurements were performed using FD-4A photodiode.
We thus measured the relative spectral transmission of the spectrograph
and determined the absolute transmission using laser emission at 0.6438~$\mu$m.
Our measurements showed that the spectrograph  optics has the transmission of
\mbox {90--95\%} practically  throughout the entire operating wavelength
range (Fig.~\ref{fig_trans}).

\subsubsection{Image Quality}

We estimated the image quality in spectroscopic mode by analyzing the
focusing dependences for series of spectra acquired with different
camera focal length values and at different positions along the slit. Our measurements showed that within the caustic, which does
not exceed 0.1~mm around the best camera focus the $FWHM$ of the slit
does not exceed 1.4~pixels.  The monochromatic size of a 1-arcsec
slit is equal to 1.22~pixels. We can thus conclude that the size of
the aberration circle of the optics in spectroscopic mode does not
exceed 0.5~pixels (13~$\mu$m). The size of the secondary spectrum in
the spectrograph does not exceed 0.1~mm.

\begin{figure*}
\centerline{ \includegraphics[scale=0.65 ]{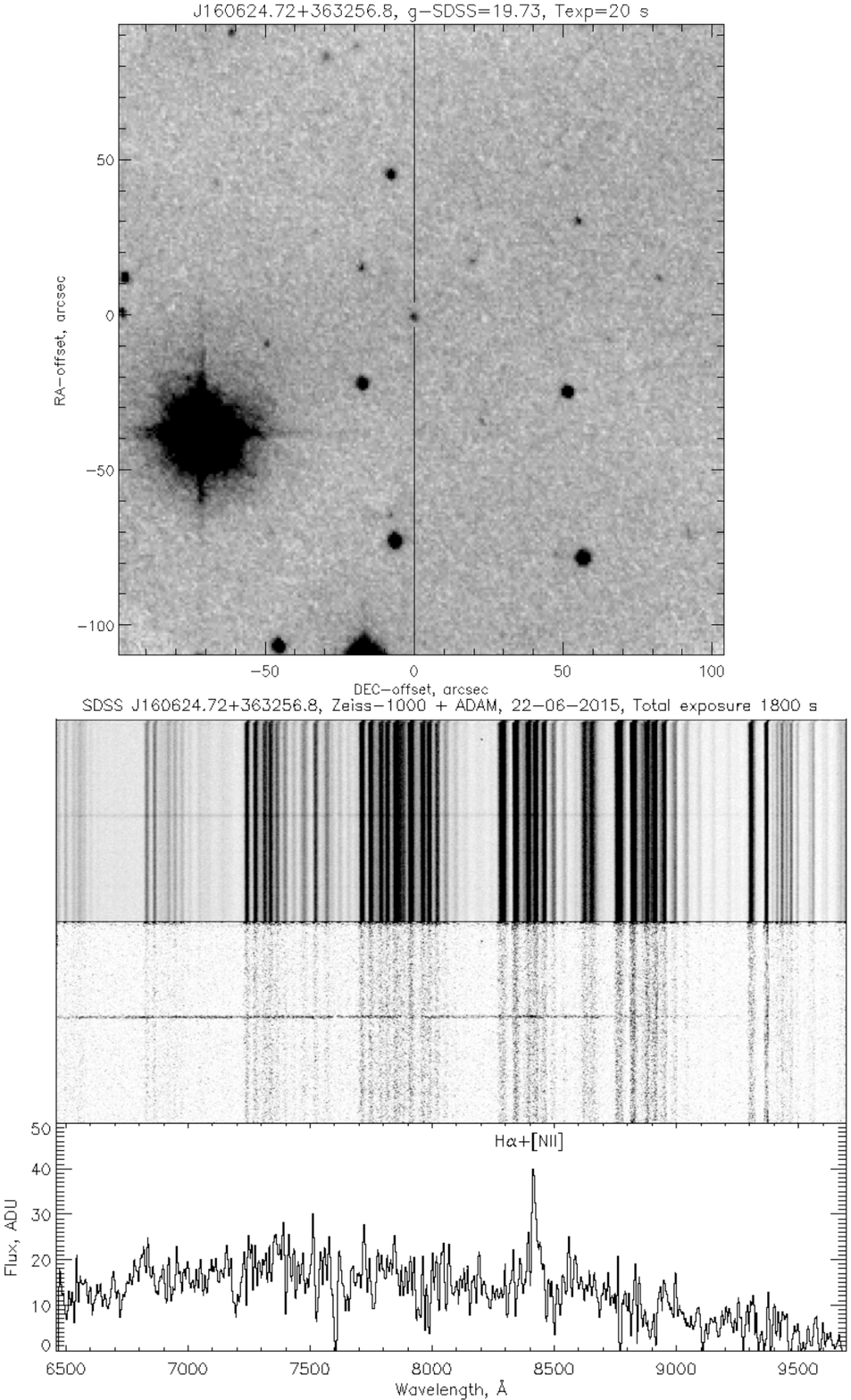} }
\caption{Observations on Zeiss-1000 telescope. The top panel shows
the direct $V$-band image of the J\,160624.72+363256.8 region. The
straight line indicates the slit position. The bottom panel shows
(from top to bottom) the 2D-spectrum, the spectrum after  subtracting
the night-sky contribution, and the integrated spectrum of the object in a
$5\arcsec$ high strip} \label{fig_spectra}
\end{figure*}

\subsubsection{Mechanical Flexure}

The amounts of flexure determined from the shift of the  image
while changing the position of the instrument on the telescope is a
crucial parameter of any astronomical  detection system.
Our measurements showed that the maximum flexure determined by
rotating the  spectrograph about the optical axis set in the horizontal position
does not exceed \mbox {$\pm10$}~$\mu$m.

\subsubsection{Thermal Conditions}

The thermal conditions of the spectrograph are guaranteed by the
thermostat mounted inside the spectrograph case and operating
independently of the control system. At ambient temperature
$-30\degr$C it takes no more than eight hours for the thermostat to
reach the working temperature of $+4\degr$C. Note that optical
surfaces and the entrance glass of the CCD do not suffer from
condensation at ambient atmospheric humidity of  75--80\%.

\subsection{Observations on the 1-m telescope}

During the period from June 17--21, 2015 test observations were
performed on the 1-m \mbox {Zeiss-1000} telescope of the SAO RAS. To
match the telescope with the spectrograph, we made a two-lens
converter reducing the aperture ratio from $F/13$ down to $F/20$.
Note that the converter fulfills the telecentrism condition---the
path of beams is equivalent to the path of rays of the Sayan
Observatory telescope, and the the size of the exit pupil
corresponds to the size of the pupil produced when ADAM spectrograph
is mounted on the 1.6-m telescope.

We  observed  spectrophotometric standards in order to determine the quantum
efficiency and acquire spectra of faint targets of various types.
Observations were carried out in the  remote mode from the SAO RAS institute building. Unfortunately, weather was bad during
our observations and only during the night of June 18/19 we could
acquire a spectrum for the standard star BD+33\,2642 of satisfactory
quality in the red to be suitable for estimating the quantum
efficiency. We estimated the extraatmospheric quantum efficiency of
the ``telescope + spectrograph + CCD'' system to be of about 60\%,
which is consistent with the expected value. During the last night of
the set, June 21/22, we acquired the spectrum of the  Seyfert
galaxy J\,160624.72+363256.8 (with the SDSS $g$\mbox{-}band magnitude
of $19\fm7$), which  has a redshift of 0.281 according SDSS data. The seeing during observations
was equal to about $2\farcs8$ and atmospheric transparency was
satisfactory. The width of the spectrograph slit was $1\farcs5$ and
ADU=2.6~$e^-$. The flux from the
target that got through the slit corresponded to an $R$-band
magnitude of about $20^{\rm m}$.
Figure~\ref{fig_spectra} shows the direct image and
reduced spectrum of the target. As is evident from the figure, an
$S/N$ ratio of about 6--7 can be achieved with a 30\mbox{-}min
exposure, demonstrating the high efficiency of the spectrograph.

\begin{figure}
\includegraphics[scale=0.55]{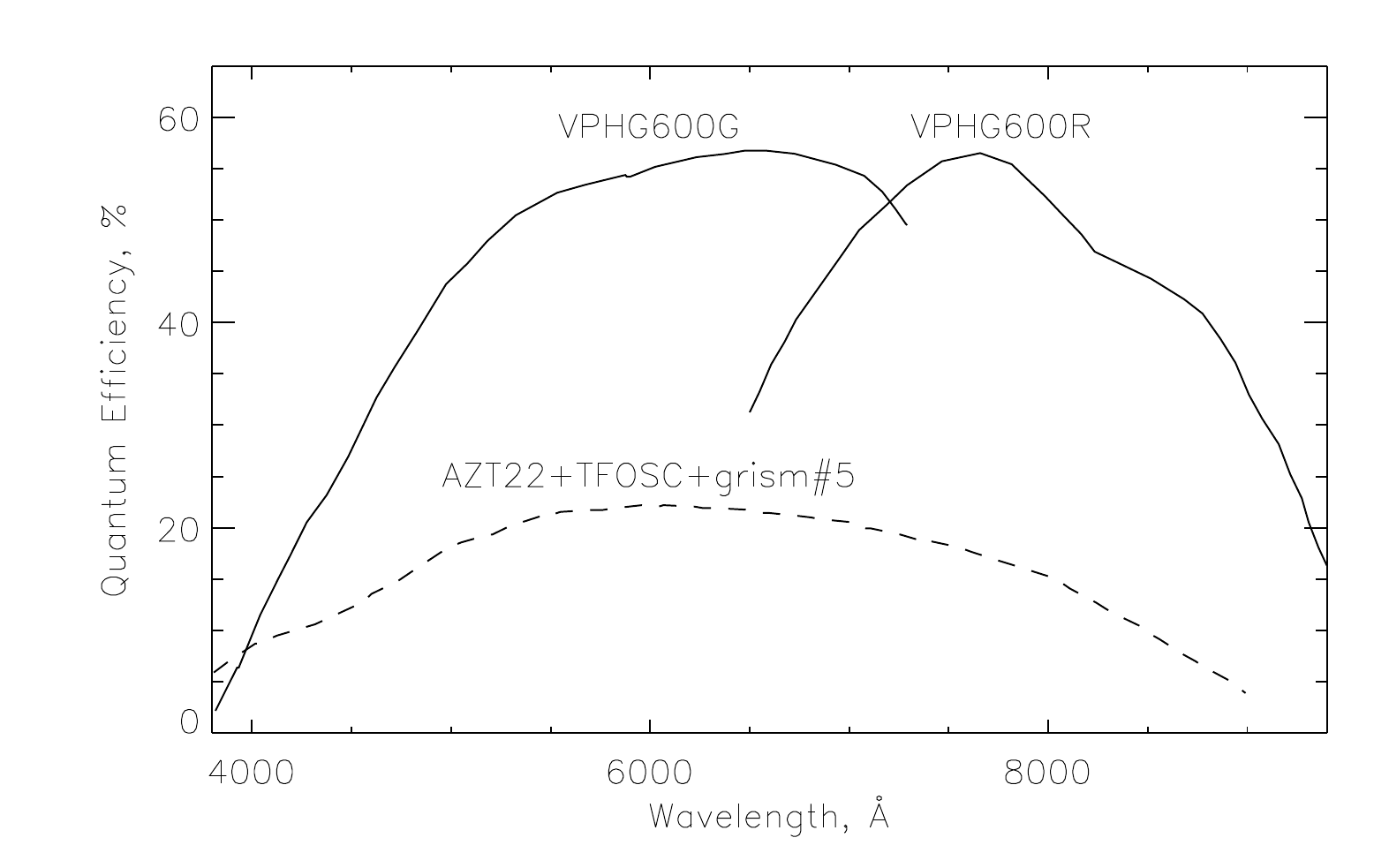} 
\caption{Curves of the total quantum efficiency of ADAM spectrograph
mounted on AZT-33IK telescope with VPHG600G and VPHG600R grisms
on the night with the best transparency (the solid lines).
The dashed line shows the similar data for  TFOSC spectrograph
of the 1.5-m RTT-150 telescope.} 
\label{fig_DQE}
\end{figure}
 
\subsection{Observations on the 1.6-m Telescope}
 
The spectrograph was mounted on the 1.6-m telescope of Sayan
Observatory in September 2015. We observed both spectrophotometric
standards and starlike targets of various types and brightness.
Transparency varied by 20--30\% from night to night.
Figure~\ref{fig_DQE} shows the curves of the total
quantum efficiency of ADAM spectrograph with AZT-33IK telescope for
the night of the best transparency. The slit width in these
observations was equal to  $10''$ for the typical seeing of \mbox
{$\theta\approx2''$}. The same figure also shows the quantum
efficiency curve for TFOSC spectrograph, which was kindly  provided
to us by R.~A.~Burenin. The spectra of various objects acquired
during these observations  are presented by
\citet{burenin}. For example, in the case of
seeing and slit width of $1\farcs5$--$2\arcsec$ and $1\farcs5$,
respectively, observing an $m_r=19.1$ elliptical galaxy with
30\mbox{-}minute exposure using VPHG600G grism yields a
signal-to\mbox{-}noise ratio as high as $S/N=10$--$15$ in the
resulting absorption spectrum. Observations with a VPHG300 grism
under the same conditions produce the spectrum of a $m_r=20.1$ quasar
with a similar signal-to-noise ratio. Furthermore, one-hour exposure
with VPHG600R produced the spectrum of a distant ($z=6.3$) quasar
with $m_r=21.0$.

\section{Conclusions}

Our experience in the development of spectroscopic instruments for the
6-m telescope of the SAO RAS combined with available modern technologies
(volume phase holographic gratings, deep depletion CCDs, industrial computers,
multilayer antireflection coatings) allowed us to make a sufficiently compact,
easy to control, and transparent optical spectrograph.
The maximum quantum efficiency of the entire system (including the telescope)
exceeds 50\%. Relatively short total exposures (of about one hour) on
AZT-33IK telescope allow acquiring spectra of starlike objects with
integrated magnitudes  $m_{\rm AB}\approx20$--$21$. This is a good result
for a 1.5-m class telescope.
 
\begin{acknowledgements}
We are grateful to S.~V.~Drabek and V.~V.~Komarov for their assistance
with the organization of test observations at the SAO RAS, T.~A.~Fatkhullin for developing
CCD control software, M.~V.~Eselevich and A.~L.~Amvrosov for their assistance
with adapting the spectrograph to AZT-33IK telescope, and to the anonymous
reviewer for the comments that improved the paper.
\end{acknowledgements}


\begin{thebibliography}{15}
 

\bibitem[{Afanasiev \& Moiseev(2005)}]{AfanasievMoiseev2005}
{Afanasiev}, V.~L., {Moiseev}, A.~V. 2005, Astronomy Letters, 31, 194

 
\bibitem[{Afanasiev \& {Moiseev}(2011)}]{AfanasievMoiseev2011}
{Afanasiev}, V.~L., {Moiseev}, A.~V. 2011, Baltic Astronomy, 20, 363


\bibitem[{Andersen et al.(1995)}]{ander}
Andersen J., Andersen M.~I., Klougart J. et~al., 1995, ESO Messenger, 79, 12  


\bibitem[{Barden et al.(2000)}]{barden}
{Barden} S.~C., {Arns} J.~A., {Colburn} W.~S., {Williams} J.~B. 2000, \pasp, 112, 809 

\bibitem[{Burenin et al.(2016)}]{burenin}
{Burenin} R.~A., {Amvrosov} A.~L.,  Eselevich M.~V., et~al. 2016, Astronomy Letters, 42, 295 

 \bibitem[{Buzzoni et al.(1984)}]{buzoni}
 {Buzzoni} B., {Delabre} B., {Dekker} H.~, et~al. 1984, ESO Messenger, 38, 9 

\bibitem[{Geyer(1975)}]{gey}
{Geyer} E.~H, 1975, Jena Review, 20, 26 

\bibitem[{Mack et al.(2010)}]{mack}
Mack P., KanniahPadmanaban S.~Y., Kaitchuck R., Borstad A., Luzier N. 2010, Bull. Amer. Astron. Soc., 41, 824 

\end{thebibliography}
\end{document}